\documentclass[aps,prb,twocolumn,superscriptaddress]{revtex4}
\usepackage{graphicx}
\usepackage{amsmath}
\usepackage{amsfonts}
\usepackage{times}

\begin{document}

\title{Edge-mode velocities and thermal coherence of quantum Hall
  interferometers}

\author{Zi-Xiang Hu}
\affiliation{Asia Pacific Center for Theoretical Physics,
Pohang, Gyeongbuk 790-784, Korea}

\author{E. H. Rezayi}
\affiliation{Department of Physics, California State University
Los Angeles, Los Angeles, California 90032, USA}

\author{Xin Wan}
\affiliation{Asia Pacific Center for Theoretical Physics,
Pohang, Gyeongbuk 790-784, Korea}
\affiliation{Department of Physics, Pohang University of Science and
Technology, Pohang, Gyeongbuk 790-784, Korea}
\affiliation{Zhejiang Institute of Modern Physics, Zhejiang
University, Hangzhou 310027, P.R. China}

\author{Kun Yang}
\affiliation{National High Magnetic Field Laboratory and
Department of Physics, Florida State University, Tallahassee,
Florida 32306, USA}

\date{\today}

\begin{abstract}

  We present comprehensive results on the edge-mode velocities in a
  quantum Hall droplet with realistic interaction and confinement at
  various filling fractions. We demonstrate that the charge-mode
  velocity scales roughly with the valence Landau level filling
  fraction and the Coulomb energy in the corresponding Landau
  level. At Landau level filling fraction $\nu = 5/2$, the stark
  difference between the bosonic charge-mode velocity and the
  fermionic neutral-mode velocity can manifest itself in the thermal
  smearing of the non-Abelian quasiparticle interference. We estimate
  the dependence of the coherence temperature on the confining
  potential strength, which may be tunable experimentally to enhance
  the non-Abelian state.

\end{abstract}

\maketitle

\section{Introduction}

Fractional quantum Hall (FQH) states are incompressible quantum
liquids that support fractionalized quasiparticles and gapless edge
excitations~\cite{wen92}. The simplest edge theory for the Laughlin
states contains a single branch of charge mode, which is described by
chiral bosons~\cite{wen90}. The velocity of the charged bosons $v_c$,
which enters the effective theory through gauge-fixing condition,
depends on electron-electron interaction and edge confinement in
general and is, therefore, not a universal quantity. Recent
development of quantum Hall interferometers~\cite{camino06,mcclure09}
also allows experimental determination of $v_c$ from current
oscillations in the devices.

Apart from the charge mode, which seems to dominate the edge tunneling
measurement,~\cite{grayson98} neutral edge modes may arise in the
hierachical states~\cite{macdonald90,kane94,hu08} or through edge
reconstruction.~\cite{chamon,wan02,yang03,wan03,overbosch08a} The
velocities of the neutral modes $v_n$, which no experiments have
observed, are conceivably smaller than $v_c$~\cite{lee98,lopez99}. At
Landau level (LL) filling fraction $\nu =
5/2$,~\cite{moore91,levin07,lee07} a neutral Majorana fermion mode
exists in the Moore-Read phase. In the non-Abelian FQH state,
numerical studies~\cite{wan06,wan08} found $v_n$ to be roughly 10
times smaller than $v_c$ in a model with long-range interaction and
confining potential due to neutralizing background charge. Even
smaller $v_n/v_c$ has been proposed to explain tunneling conductance
measurement at $\nu = 2/5$.~\cite{ferraro08}

The stark difference in the edge-mode velocities in the Moore-Read
phase can lead to ``Bose-Fermi separation'',~\cite{wan08} which
resembles spin-charge separation in Luttinger liquids. When
propagating along the edge, quasiparticles with charge $e/4$ carrying
both charged bosonic and neutral fermionic components smear at finite
temperatures. This poses a stringent requirement on the observation of
current oscillations due to quasiparticle interference in a double
point-contact
interferometer,~\cite{chamon97,fradkin98,dassarma05,stern06,bonderson06}
whose contact distance should not exceed the quasiparticle dephasing
length.~\cite{bishara08} On the other hand, the Abelian charge $e/2$
quasiparticles carrying the bosonic component {\em only} (and thus not
affected by the slow fermionic mode) smear significantly less and may
dominate the current oscillations at higher temperatures when charge
$e/4$ quasiparticle transport is incoherent,~\cite{wan08} even though
the tunneling amplitude of the charge $e/2$ quasiparticles is much
smaller than that of the charge $e/4$
quasiparticles.~\cite{chen09,bishara09a}

Recently, experimentalists observed~\cite{willett09} conductance
oscillations consistent with the interference of both $e/4$ and $e/2$
quasiparticles in a double point contact interferometer at $\nu=5/2$,
and the dominance of $e/2$ periods at higher temperatures when the
conductance oscillations are still visible.  Further observation of
the alternative $e/4$ and $e/2$ oscillations~\cite{willett} is now
under examinations~\cite{bishara09a} on its potential origin in the
odd-even effect~\cite{stern06,bonderson06} due to non-Abelian
statistics, mingled with the bulk-edge coupling
effect.~\cite{rosenow08,rosenow09,bishara09b} The interpretation of
the experimental findings as non-Abelian
interference~\cite{bishara09a} (as opposed to, e.g., the charging
effect~\cite{rosenow07,ihnatsenka08,zhang09}) relies crucially on the
estimate of edge-mode velocities from numerical studies,~\cite{wan08}
which predicted significant difference in coherent lengths or coherent
temperatures for charge $e/4$ and $e/2$ quasiparticles.

In this paper, we report results on studies of edge-mode velocities in
a realistic microscopic model for both Abelian and non-Abelian FQH
liquids, including the analysis on finite-size effect and the effect
of edge confinement. We show that the charge-mode velocity $v_c$ is
roughly proportional to the valence LL filling fraction and the
Coulomb energy scale in the corresponding LL, while the neutral-mode
velocity $v_n$ in the Moore-Read phase is consistently smaller.  We
further present our estimate of the coherent temperatures for charge
$e/4$ and $e/2$ quasiparticles in the Moore-Read phase based on the
quantitative knowledge of the two velocities, and show that they are
in quantitative agreement with recent experiments.

The rest of the paper is organized as follows.  We introduce our model
and discuss its relevance to an experimental setup in
Sec.~\ref{sec:model}. We present our main results on edge-mode
velocities in Sec.~\ref{sec:results} and discuss their implication on
interference experiments in Sec.~\ref{sec:interference}. We discuss
the limitation of our microscopic approach and summarize the paper in
Sec.~\ref{sec:discussion}.

\section{Groundstatability in a charge-neutral model}
\label{sec:model}

We consider a microscopic model of a two-dimensional electron gas
(2DEG) in a disk geometry with long-range Coulomb interaction among
electrons in the lowest Landau level (0LL) or the first excited Landau
level (1LL). We introduce a confining potential by including
neutralizing background charge distributed uniformly on a parallel
disk at a distance $d$ from the 2DEG.  The model prototypes a
modulation-doped sample, which separates charged impurities spatially
from the electron layer for high charge mobility. We use the location
of the background charge to tune from stronger confinement (smaller
$d$) to weaker confinement (larger $d$). The microscopic Hamiltonian
is
\begin{equation}
\label{eqn:chamiltonian} H_{\rm C} = {1\over 2}\sum_{mnl}V_{mn}^l
c_{m+l}^\dagger c_n^\dagger c_{n+l}c_m +\sum_m U_mc_m^\dagger c_m,
\end{equation}
where $c_m^\dagger$ creates an electron in a single-particle state
with angular momentum $m$ in the symmetric gauge. $V_{mn}^l$'s are the
corresponding matrix elements of Coulomb interaction, and $U_m$'s the
matrix elements of the confining potential.~\cite{wan03,wan08} The
definitions of the electron operators and matrix elements are
consistent with the corresponding LL index; the common practice is to
map them into the 0LL representation using ladder operators across
LLs.~\cite{macdonald84} At various $\nu$, we study the spectrum of the
model Hamiltonian numerically by exact diagonalization using the
Lanczos algorithm up to 7 electrons for $\nu = 1/5$, 10 electrons for
$\nu = 1/3$, 14 electrons for $\nu = 5/2$, and 20 electrons for $\nu =
2/3$.

\begin{figure}
\includegraphics[width=8cm]{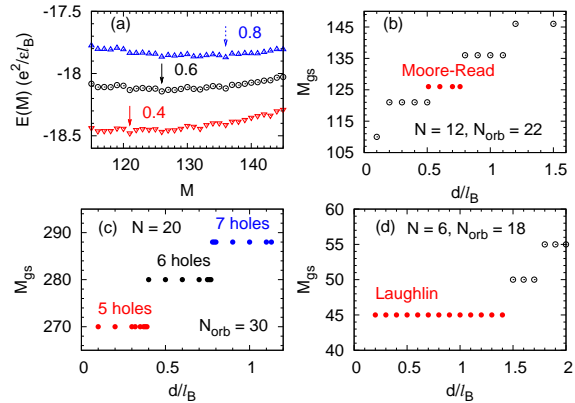}
\caption{\label{fig:gs} (color online) (a) Ground state energy $E(M)$
  in each total angular momentum $M$ subspace for $N = 12$ electrons
  with 22 orbitals in the 1LL ($\nu = 5/2$). The neutralizing
  background charge is at a distance $d/l_B = 0.4$, 0.6, and 0.8 away
  from the 2DEG. The total angular momentum of the global ground state
  $M_{\rm gs}$ (indicated by arrows) increases from 121, 126, to 136,
  respectively. The global ground state at $d = 0.6 l_B$ has the same
  total angular momentum $M_{\rm MR} = N(2N-3)/2 = 126$ as the
  corresponding $N = 12$ Moore-Read state. We have shifted the curves
  verically for $d = 0.4 l_B$ and $0.8 l_B$ for clarity. (b) $M_{\rm
    gs}$ as a function of $d$ at $\nu = 5/2$. The solid dots indicate
  the plateau at which the Moore-Read state is groundstatable.  (c)
  $M_{\rm gs}$ as a function of $d$ for 20 electrons in 30 orbitals
  ($\nu = 2/3$). We can understand the three plateaus of $M_{\rm gs}$
  as $\nu = 1/3$ Laughlin droplets of 5-7 holes embedded in a $\nu =
  1$ electron droplet. Note that they are in the same phase, but the
  corresponding distances between two counterpropagating edges are
  different.~\cite{hu08} (d) $M_{\rm gs}$ as a function of $d$ for 6
  electrons in 18 orbitals ($\nu = 1/3$). The solid dots indicate the
  plateau at which the Laughlin state is groundstatable. Beyond $d
  \approx 1.5 l_B$, the system undergoes an edge-reconstruction
  transition.~\cite{wan02,wan03} }
\end{figure}

Our approach features the inclusion of a realistic confining potential
due to the neutralizing background charge; it allows us to study the
energetics in the microscopic model quantitatively and compare with
the experimental observations. For example, we can identify various
phases by studying the total angular momentum of the ground state
$M_{\rm gs}$ of the model as $d$ varies at a given $\nu$, as
illustrated in Fig.~\ref{fig:gs}. In Fig.~\ref{fig:gs}(a), we plot the
ground state energy $E(M)$ in each total angular momentum ($M$)
subspace for $N = 12$ electrons in 22 1LL orbitals, i.e. at $\nu =
5/2$. When the background charge lies at $d/l_B = 0.4$, 0.6, and 0.8,
$M_{\rm gs}$ yields 121, 126, and 136, respectively. The global ground
state at $d = 0.6 l_B$ has the same $M_{\rm gs}$ as that of the $N =
12$ Moore-Read state $M_{\rm MR} = N(2N-3)/2 = 126$; thus we say that
the Moore-Read state is {\it groundstatable}~\cite{charrier08} at $d =
0.6 l_B$. Our calculation shows that the 12-electron Moore-Read state
is groundstatable in 22 1LL orbitals with $0.51 \leq (d/l_B) \leq
0.76$, as indicated in Fig.~\ref{fig:gs}(b). We have shown that the
ground state evolves continuously into the Moore-Read state when the
Coulomb interaction changes smoothly to a three-body interaction in
Ref.~\onlinecite{wan08}, which also discussed the nature of the other
ground states.

Our main motivation to introduce a neutralizing charge background at a
distance $d$ away from the 2DEG is for modeling experimental
structures. In ultra-high-mobility GaAs samples,~\cite{pfeiffer03} an
undoped setback, whose thickness optimizes charge mobility, separates
a thin layer of Si impurities and the 2DEG. Therefore, it is tempting
to identify $d$ in our model as (or to quantitatively relate it to)
the thickness of the setback layer, which can be of the order of 1000
\AA,~\cite{pfeiffer89} or about 10 magnetic length $l_B$ for a typical
magnetic field strength ($\sim$5 T, as in
Ref.~\onlinecite{willett09}). However, the direct identification is an
oversimplification. In past studies, we have encountered at least
three different scenarios in which additional edges are present in the
system and spoil the identification by changing the effective edge
confining potential. First, as in the $\nu = 5/2$ case,~\cite{wan08}
filled LLs introduce integer edges, which may influence the confining
potential of the inner edge for the partially filled LL. Second, in a
nonchiral case like $\nu = 2/3$,~\cite{hu08} two counterpropagating
edges can move relatively to each other to adjust their shares of the
confining potential, leading to a series of ground states with
different $M_{\rm gs}$, as illustrated in
Fig.~\ref{fig:gs}(c). Nevertheless, they belong to the same
phase.~\cite{hu08} Finally, even in the case of a single chiral edge,
a fractional quantum Hall droplet may undergo one or more
edge-reconstruction transitions to introduce counterpropagating edges
for $d > 1.5 l_B$ [as illustrated in Fig.~\ref{fig:gs}(d)], beyond
which the microscopic calculation suffers from size
limitation.~\cite{wan03} However, even with the complications, we
believe that the introduction of $d$ is experimentally relevant in the
following two ways. First, the range of $d$ in which a specific phase
is groundstatable often indicates the stability of the phase, as the
comparison of Figs.~\ref{fig:gs}(b) and \ref{fig:gs}(d) clearly
demonstrates. Second, the relative value of $d$ measures the relative
strength of the confining potential, thus allowing qualitative and,
sometimes, quantitative predictions from the model calculation.

\section{Edge-mode velocities}
\label{sec:results}

\begin{figure}
\includegraphics[width=8cm]{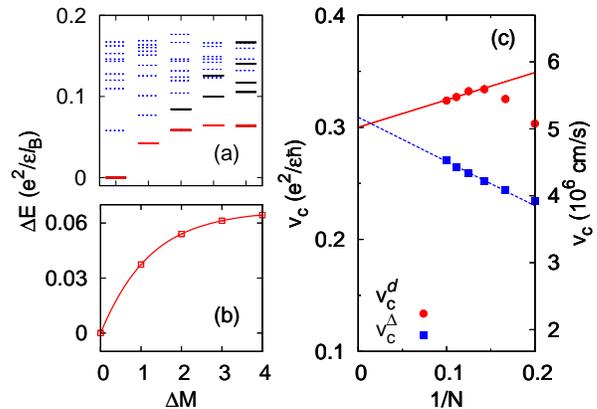}
\caption{\label{fig1} (color online) Edge-mode velocity analysis for
  $\nu = 1/3$.  (a) Low-energy excitations in a $N = 8$ system with
  background charge at $d = 0.6l_B$, with edge excitations marked by
  solid bars. (b) Energies of single chiral boson excitations, which
  are the ground states in individual momentum spaces [marked in red
    in (a)]. (c) Chiral boson (or charge mode) velocity $v_c$ defined
  either by the slope of the best quadratic fit in (b) at the origin
  ($v_c^d$) or by the ground-state energy difference at $\Delta M = 0$
  and 1 ($v_c^{\Delta}$) for systems of 5-10 electrons. The
  finite-size scaling of $v_c^{\Delta}$ shows a weaker nonlinear
  effect, although the two definitions tend to give the same result in
  the thermodynamic limit.}
\end{figure}

\begin{figure}
\includegraphics[width=8cm]{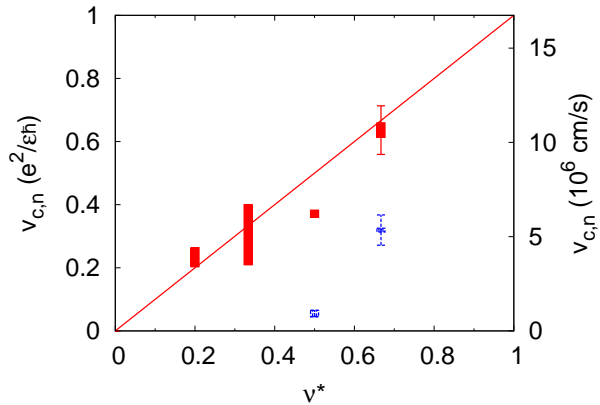}
\caption{\label{fig2} (color online) Summary of the charge-mode
  velocity $v_c$ (solid red candles) and neutral-mode velocity $v_n$
  (empty blue candles), extrapolated to the thermodynamic limit
  (except for $\nu = 2/3$), at various valence LL filling $\nu^{*} =
  \nu - \lfloor \nu \rfloor$. The size of the candles reflects the
  range of velocities in the corresonding groundstatable range of the
  confining potential strength (or $d$). The candle data for $\nu =
  2/3$ are based on 20 electrons in 30 orbitals for various confining
  potential strength, while the errorbars reflect our estimate of the
  uncertainty due to finite-size effects. The solid line is the
  dimensional analysis result $v_c = \nu^{*} e^2 / \epsilon \hbar$.}
\end{figure}

At $\nu = 1/3$, the edge excitations arise from a single branch of
bosonic charge mode as illustrated in Fig.~\ref{fig1}(a) for $N = 8$
electrons and $d = 0.6 l_B$. Ref.~\onlinecite{wan03}, in the context
of edge reconstruction, has explained in detail the method for the
identification of the edge excitations among all excitations. In a
finite system, we can then define $v_c$ through (i) the excitation
energy $\Delta E (\Delta M = 1)$ of the smallest momentum mode with
edge momentum $k = \Delta M/R = 1/R$, i.e. $v^{\Delta}_c = (R / \hbar)
\Delta E(\Delta M = 1)$, where $R = \sqrt{2N/\nu}$ is the radius of
the droplet, or (ii) the slope at $k = 0$ of the dispersion curve [see
  Fig.~\ref{fig1}(b)] fitted to the edge-mode energies, i.e. $v^{d}_c
= (R/\hbar) (dE/dM)$. As shown in Fig.~\ref{fig1}(c), both definitions
point to $\sim$$0.3 e^2/\epsilon \hbar$ in the thermodynamic limit and
are therefore suitable. The definition $v^{d}_c$ is less robust due
to the nonlinear dispersion at small $k$ in finite systems; thus, we
will use the definition $v^{\Delta}_c$ in later analyses. In general,
$v_c$ depends on the strength of the confining potential, which is
determined by {\it charge neutrality} and is meaningful only when the
Laughlin variational wavefunction is {\it
  groundstatable},~\cite{charrier08} i.e. the Coulomb ground state has
the same angular momentum as the Laughlin state. For $\nu=1/3$,
neutrality limits $v_c$ from above, while groundstatability from
below,~\cite{wan02,wan03} allowing $v_c$ to vary between
$0.21$$\sim$$0.40 e^2/\epsilon \hbar$ (see Fig.~\ref{fig2}), or
$3.5$$\sim$$6.7 \times 10^6$ cm/s in GaAs systems (with dielectric
constant $\epsilon \approx 13$).  A similar analysis for $\nu = 1/5$
(with less robust finite-size analysis due to computational
limitation) reveals $v_c$ to be $0.20$$\sim$$0.26 e^2/\epsilon \hbar$,
or $3.4$$\sim$$4.4 \times 10^6$ cm/s for GaAs.

We now turn to $\nu = 5/2$, or a valence LL filling $\nu^{*} = \nu -
\lfloor \nu \rfloor = 1/2$, after subtracting the filled 0LL.  The
even-denominator quantum Hall state is widely expected to be the
Moore-Read Pfaffian state,~\cite{moore91} described by the SU(2)$_2$
topological quantum field theory, or its particle-hole conjugate, the
anti-Pfaffian state.~\cite{levin07,lee07} In numerical studies, the
Pfaffian state can be stabilized on the disk
geometry,~\cite{wan06,wan08} whose edge contains both neutral Majorana
fermions and charged chiral bosons. The heavy mixture of the bulk and
edge excitations complicates the extraction of $v_{c,n}$.~\cite{wan06}
Therefore, we mix the 3-body interaction with the Coulomb interation
to separate the edge and bulk excitations in energy and extrapolate
the calculated velocities to the pure Coulomb case.~\cite{wan08} Based
on a similar finite-size scaling in systems of 6-14 electrons as
described above, we obtain $0.36 e^2/\epsilon \hbar < v_c < 0.38
e^2/\epsilon \hbar$, or $6 \times 10^6 < v_c < 6.4 \times 10^6$ cm/s
for GaAs, which is about 20\% larger than the estimate based on a
12-electron system.~\cite{wan08}

We summarize the range of $v_c$ in Fig.~\ref{fig2}. For $\nu = 2/3$,
we include both the left- and right-going-mode
velocities,~\cite{twothird}~\cite{hu08} arising from the coupling of
the edge modes~\cite{wen92,kane94} of the inner Laughlin $\nu = 1/3$
hole droplet and the outer $\nu = 1$
droplet~\cite{macdonald90,johnson91}. Since the Coulomb energy is the
only energy scale, dimensional analysis implies 
\begin{equation}
\label{eq:vc}
v_c \sim \nu^{*} e^2
/ \epsilon \hbar = (\nu^{*} \alpha / \epsilon)c, 
\end{equation}
where $c$ is the speed of light in vacuum and $\alpha \approx 1/137$
the fine-structure constant. Our numerical results, including the
larger velocity for $\nu = 2/3$ (which we assume to be the charge-mode
velocity), agrees well with Eq.~(\ref{eq:vc}).  At $\nu = 5/2$, we can
attribute the deviation to the reduction of the Coulomb energy scale
in the 1LL, due to the change of the LL structure factor.  On the
other hand, the counterpropagating (presumably neutral) mode at $\nu =
2/3$ has a velocity closer to the value for $\nu = 1/3$; in finite
systems, we cannot resolve whether the interedge Coulomb coupling
(divergent in the long wavelength limit) can lower the value (as being
decoupled from charge, the neutral mode has a conceivably small
velocity). Recently, trial wavefunctions are under consideration for
such FQH states with negative flux in the composite fermion
approach.~\cite{milovanovic08} It is worth mentioning that the results
of $v_c$ are consistent with the recent experimental determinations of
$v_c$ to be $4 \times 10^6$ cm/s for $\nu = 1/3$~\cite{camino06} or to
saturate at $1.5 \times 10^7$ cm/s in the integer
regime,~\cite{mcclure09} further justifying the validity of our
microscopic model calculation.  They are also quantitatively
consistent with earlier time-resolved transport
measurements.~\cite{ashoori92,zhitenev93,note}

On the other hand, theoretical proposals agree on a smaller value for
$v_n$, with electrostatic,~\cite{lee98} topological,~\cite{lopez99}
and dynamic origins.~\cite{yu07} But the edge magnetoplasmon
experiment in the time domain,~\cite{ashoori92} e.g., showed no hint
of the existence of a counterpropagating edge mode at $\nu = 2/3$,
although theories~\cite{macdonald90,johnson91,wen92,kane94} and the
numerical study on a realistic model~\cite{hu08} suggested its
existence. In the chiral case at $\nu = 5/2$, we found $v_n$ to be one
order of magnitude smaller than $v_c$ in a system with $N = 12$
electrons at $d = 0.6 l_B$.~\cite{wan08} Here, we extend the study on
$v_n$ by exploring the parameter space of $d$ and $N$, as long as the
Moore-Read phase is the stable ground state, or {\it groundstatable},
with Coulomb interaction among electrons. As explained in earlier
studies,~\cite{wan08} we first mix 3-body interaction with Coulomb
interaction to allow a clear separation of bulk modes and edge modes.
In this case, the lowest edge excitation with $\Delta M = 2$ has an
excitation energy $\Delta E_n$ being the sum of the two neutral
excitation energies at $kR = 1/2$ and $3/2$ due to the anti-periodic
boundary condition in the absence of bulk quasiparticles. Therefore,
we expect $v_n = (R / 2 \hbar) \Delta E_n (\Delta M = 2)$.
Extrapolating data to the pure Coulomb case and to the thermodynamic
limit, we obtain the range of $v_n$ to be $0.045$$\sim$$0.065
e^2/\epsilon \hbar$, or $0.75$$\sim$$1.1 \times 10^6$ cm/s for
GaAs. The finite-size scaling analysis reveals that the value of $v_n$
in a 12-electron system obtained earlier~\cite{wan08} is about 40\%
smaller than its thermodynamic value, though we still have a large
ratio $v_c/v_n \approx 6$$\sim$8.

Within the groundstatable range of the parameter space, the velocities
decrease as $d$ increases, i.e., as the confining potential strength
becomes weaker. We can fit the trend by a linear dependence on $d$. At
$\nu = 5/2$, we find that the velocities in the thermodynamic limit are 
\begin{eqnarray}
\label{eq:trend1}
v_c &=& 0.435 - 0.106 d / l_B \\
\label{eq:trend2}
v_n &=& 0.123 - 0.120 d / l_B,
\end{eqnarray}
in unit of $e^2/\epsilon \hbar$.

\section{Thermal decoherence in interference experiments}
\label{sec:interference}

\begin{figure}
\includegraphics[width=8cm]{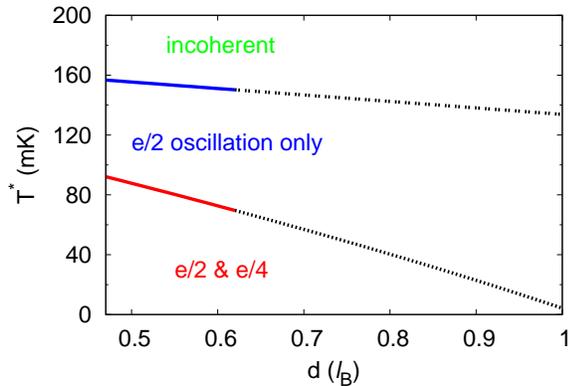}
\caption{\label{fig3} (color online) Coherence temperature $T^{*}$
  (based on the estimate of $v_c$ and $v_n$) as a function of $d$ for
  both e/4 (upper line) and e/2 (lower line) quasipaticles in the
  Moore-Read state in an $L = 1$ $\mu$m sample. The broken lines at $d
  > 0.62 l_B$ are obtained by the extrapolation of the velocities, as
  the Moore-Read state is no longer groundstatable in a system of 12
  electrons in 26 orbitals. A stripe phase may emerge below $d =
  l_B$.~\cite{wan08}}
\end{figure}

The small $v_n$, as opposed to the larger $v_c$, dominates the thermal
smearing of the non-Abelian $e/4$ quasiparticles in edge transport.
In an interference experiment, finite temperature $T$ introduces an
additional energy scale to compete with the traverse frequency for
quasiparticles (or edge waves) propagating from one point contact to
another between multiple reflections within the interferometer.
Therefore, the Aharonov-Bohm type oscillation will be washed out above
the coherence temperature~\cite{bishara09a}
\begin{equation}
T^* = \frac{1}{2 \pi L k_B} \left( \frac{g_c}{v_c} +
\frac{g_n}{v_n} \right )^{-1},
\end{equation}
where $L$ is the distance along the interference path between the two
point contacts. For the Moore-Read Pfaffian (or anti-Pfaffian) state,
$g_c = 1/8$ and $g_n = 1/8$ (or $3/8$) are the corresponding scaling
dimensions. Using our results for $v_c$ and $v_n$, we plot $T^{*}$ in
the Moore-Read phase for charge $e/4$ quasiparticles and the $e/2$
quasiparticles in the Ising vacuum sector ($g_c = 1/2$ and $g_n = 0$)
as a function of $d$ in Fig.~\ref{fig3}, assuming a point contact
distance of $L = 1$ $\mu$m. Here, we use the extrapolated value of
$v_c$ and $v_n$ in the thermodynamic limit for the Coulomb interaction
[Eqs.~(\ref{eq:trend1}) and (\ref{eq:trend2})]. Therefore, in an
interference experiment, we expect that with increasing $T$ the
conductance oscillations due to the interference of $e/4$
quasiparticles (thus signatures of non-Abelian statistics) will
disappear first at a lower temperature, which depends strongly on the
details of the confinement potential. On the other hand, the
oscillations due to $e/2$ quasiparticles will persist up to about 150
mK, which is less sensitive to the confinement. This picture and the
corresponding temperature ranges agree with the Willett
experiment~\cite{willett09} quantitatively.

For the solid range of the curves in Fig.~\ref{fig3} (or $0.47 \le
d/l_B \le 0.62$), we can justify the groundstatability of the
Moore-Read state in a 12-electron system in 26 orbitals with the pure
Coulomb interaction. This range is sensitive to the system size and
the number of orbitals; e.g., we found it to be $0.51 \le d/l_B \le
0.76$ for the 12-electron system in 22 orbitals. This indicates that
the stability of the Moore-Read state is sensitive to the sharpness of
the edge confinement. In general, we find the groundstatable range
becomes wider in larger systems; therefore, it is tempting to
extrapolate the trend in velocities [Eqs.~(\ref{eq:trend1}) and
(\ref{eq:trend2})] to $d_c \approx l_B$, where $v_n$ vanishes.
However, a potentially competing stripe phase may arise below
$d_c$.~\cite{wan08} Accordingly, we extrapolate the coherence
temperatures further in weaker confinement, as illustrated by the
broken lines in Fig.~\ref{fig3}.

One can probe the sensitive dependence of $T^{*}$ on $d$ for $e/4$
quasiparticles by controlling $l_B$ via tuning electron density and
magnetic field simultaneously or by applying an external confining
potential. This would indirectly support our conclusion that it is the
smallness of $v_n$ that controls the coherence length or
temperature. Direct evidence may come from the transport measurements
of a long tunneling contact~\cite{overbosch08b} or from the
momentum-resolved tunneling measurements,~\cite{seidel09,wang09} in
which the slower neutral mode is accessible.

\section{Discussion and conclusion}
\label{sec:discussion}

The narrow range of both charge-mode and neutral-mode velocities,
Abelian or not, appears to be surprising, as the velocities are
sensitive to confining potential and electron-electron
interaction. This is, however, understandable.  Despite the stability
in the bulk, edges of fractional quantum Hall states are
fragile.\cite{wan02,wan03} For $\nu=5/2$, multiple competing phases
also exist in the bulk.~\cite{levin07,lee07,blok92,halperin83,wan08}
Therefore, a significant change of the edge-mode velocities must
follow from significant changes of parameters, which normally lead to
various instabilities. For example, when background confining
potential becomes weaker, edge
reconstruction~\cite{wan02,yang03,wan03} takes place, changing the
ground state and creating additional neutral modes with smaller
velocities. Consequently, $v_c$ increases in the reconstructed case
(see Fig. 10 in Ref.~\onlinecite{wan03}). It is, however, possible
that $v_n$ can become smaller in the weaker confining potential that
leads to edge reconstruction.~\cite{ferraro08}

In reality, the finite spread of the electron wavefunction in the
perpendicular direction tends to soften the Coulomb interaction.
While we focus strictly on zero layer thickness in the presentation,
the velocity ranges cannot change too much because the finite
thickness also changes the confining potential arising from charge
neutrality, so the groundstatability windows shift
accordingly.~\cite{wan03} Perhaps the most uncertain factor is the
filled lowest Landau level neglected completely in the calculation for
$\nu=5/2$.  However, we expect that the enforcement of the criterion
of groundstatability is likely to protect the velocities from
deviating significantly.

In conclusion, we have calculated the ranges of edge-mode velocities
in various FQH states using a microscopic model with long-range
interaction and tunable confining potential. In the Moore-Read phase
for $\nu = 5/2$, our calculations conclude that the charge-mode
velocity is consistently much greater than the neutral-mode velocity,
leading to the Bose-Fermi separation and the dominance of $e/2$ and
$e/4$ oscillations at different temperature ranges in a quantum Hall
interferometer,~\cite{wan08} which is consistent with the recent
interferometry experiments.~\cite{willett09,willett}

\section*{ACKNOWLEDGMENTS}

We thank Claudio Chamon and Bas Overbosch for helpful discussion on
the measurement of edge-mode velocities.  This work was supported in
part by NSF grants No. DMR-0704133 (K.Y.)  and DMR-0606566 (E.H.R.),
as well as PCSIRT Project No. IRT0754 (X.W.). X.W. acknowledges the
Max Planck Society and the Korea Ministry of Education, Science and
Technology for the joint support of the Independent Junior Research
Group at the Asia Pacific Center for Theoretical Physics.


\begin{thebibliography}{99}
\bibitem{twothird} The ground state angular momentum for $\nu = 2/3$
  depends on both $d$ and $N$ so the scaling analysis is impossible for
  accessible system sizes. We artificially introduced the finite-size
  effect (marked by the error bars in Fig.~\ref{fig2}) based on the
  amount observed in the Laughlin case.
\end{thebibliography}
\end{document}